%% file: main.tex
\documentclass[sigconf, nonacm, pdfa]{acmart}

\usepackage{pvldb}

\input{packages}
\input{macros}

\renewcommand\vldbdoi{XX.XX/XXX.XX}

\renewcommand\vldbavailabilityurl{https://github.com/snowflake-labs/avalanchebench} 

\begin{document}
\title{AvalancheBench: Evaluating Enterprise Agent-Driven Semantic Query Processing Through Latent World Recovery}

\author{Darek K{\l}eczek}
\affiliation{%
  \institution{Snowflake Inc.}
  \city{}
  \country{}
}
\email{darek.kleczek@snowflake.com}

\author{Alexander W. Lee}
\affiliation{%
  \institution{Brown University and Snowflake Inc.}
  \city{}
  \country{}
}
\email{alexander\_w\_lee@brown.edu}

\author{Julien Tissier}
\affiliation{%
  \institution{Snowflake Inc.}
  \city{}
  \country{}
}
\email{julien.tissier@snowflake.com}

\author{Fuheng Zhao}
\authornote{Work done while at Snowflake.}
\affiliation{%
  \institution{University of Utah}
  \city{}
  \country{}
}
\email{fuheng.zhao@utah.edu}

\author{Pawe{\l} Liskowski}
\affiliation{%
  \institution{Snowflake Inc.}
  \city{}
  \country{}
}
\email{pawel.liskowski@snowflake.com}

\author{U\u{g}ur \c{C}etintemel}
\affiliation{%
  \institution{Brown University and Snowflake Inc.}
  \city{}
  \country{}
}
\email{ugur\_cetintemel@brown.edu}

\author{Anupam Datta}
\affiliation{%
  \institution{Snowflake Inc.}
  \city{}
  \country{}
}
\email{anupam.datta@snowflake.com}

\input{abstract}

\maketitle

\vldbtopmatter

\input{introduction}
\input{benchmark}
\input{design}
\input{evaluation}
\input{related}
\input{conclusions}
\input{acks}

\balance

\bibliographystyle{ACM-Reference-Format}
\bibliography{main}

\end{document}

%% file: packages.tex
\usepackage{enumitem}
\setlist[itemize]{leftmargin=*, nosep}
\setlist[enumerate]{leftmargin=*, nosep}

%% file: macros.tex
\newcommand{\N}{\mathbb{N}} 
\newcommand{\R}{\mathbb{R}} 
\newcommand{\anchor}{\model_0} 
\newcommand{\arenascore}[1]{S_{#1}} 
\newcommand{\btcoef}[1]{\beta_{#1}} 
\newcommand{\capfactor}[1]{k_{#1}} 
\newcommand{\invocations}[1]{V_{#1}} 
\newcommand{\invocation}{v} 
\newcommand{\modelset}{M} 
\newcommand{\model}{m} 
\newcommand{\tokcinweight}{\tokweightsym^{\text{cin}}} 
\newcommand{\tokcin}[1]{\toksym_{#1}^{\text{cin}}} 
\newcommand{\tokcwriteweight}{\tokweightsym^{\text{cwrite}}} 
\newcommand{\tokcwrite}[1]{\toksym_{#1}^{\text{cwrite}}} 
\newcommand{\tokinweight}{\tokweightsym^{\text{in}}} 
\newcommand{\tokoutweight}{\tokweightsym^{\text{out}}} 
\newcommand{\tokout}[1]{\toksym_{#1}^{\text{out}}} 
\newcommand{\toksin}[1]{\toksym_{#1}^{\text{in}}} 
\newcommand{\toksym}{t} 
\newcommand{\tokvec}[1]{\mathbf{\toksym}_{#1}} 
\newcommand{\tokweightsym}{w} 
\newcommand{\tokweights}{\mathbf{\tokweightsym}} 
\newcommand{\world}{Z} 
\newcommand{\data}{X} 
\newcommand{\questions}{Q} 
\newcommand{\question}{q} 
\newcommand{\answers}{A} 
\newcommand{\report}{R} 

%% file: abstract.tex
\begin{abstract}
As LLMs improve, data agents are emerging as a new interface for
analytics, while semantic query processing engines use LLMs to
operate over structured and unstructured data. Existing benchmarks
largely evaluate agents and semantic query engines in isolation,
leaving it unclear whether agents can decide when and how to use
semantic operators in end-to-end analysis.

We introduce \emph{AvalancheBench}, a multimodal enterprise analytics
benchmark that generates observed data from latent world
specifications and evaluates data agents on their ability to recover that
world. The benchmark's three validated tasks contain 58 questions spanning text,
images, and documents. Questions are provided together as a multi-part analytical brief
and may share intermediate results, exposing the \emph{avalanche effect}:
errors can propagate through legitimate dependencies, and
inappropriate reuse can cause otherwise unrelated questions to fail.
AvalancheBench measures analytical fidelity, monetary cost, and \emph{capability-adjusted token cost (CATC)}, a new metric that weights token use by model capability to separate technical efficiency from provider pricing.

We evaluate five model--harness configurations across three agent
harnesses using Snowflake Cortex AISQL as the engine, with the best configuration
reaching a macro analytical fidelity of 0.82. Agents fail primarily
in constructing the analysis: failing to recover facts that require
semantic computation costs 45.1\% of lost points, yet incorrect operator outputs account for only 2.5\%, versus 36.3\% from agents misusing or avoiding semantic operators.
Misinterpreting requirements costs the most, at 51.5\%.
Errors also spread between questions: 29\% of lost points come from reusing another question's results, and avalanches including their
roots involve 42.2\% of all lost points. Finally, monetary cost and
CATC produce different Pareto frontiers, distinguishing economically
attractive configurations from those that are technically
resource efficient.
\end{abstract}

%% file: introduction.tex
\section{Introduction}
\label{sec:intro}

As LLM capabilities have continued to improve over the past few years, \emph{data agents} are emerging as a new interface for analytics.
Given a natural language question, a data agent analyzes the user's data using tools like SQL or Python and produces an answer.
Concurrently, \emph{semantic query processing engines}~\cite{cortexaisql2026,docetl2025,moar2026,lotus2025,palimpzest2025,abacus2026,caesura2023,vectraflow2025,aryn2025,continuousprompts2025,stretto2026,samsara2025,docdb2025,aop2025,binder2023,databricks_ai_functions,duckdb_prompt,bigquery_ai_functions,uqe2024} have been developed to process both structured and unstructured data using LLMs.
Queries are specified in a declarative language that composes both standard relational operators and semantic operators.
Agents can leverage these semantic query processing engines as tools, but given the recency of their development, it is unclear how effectively agents use them, especially in real enterprise settings.

Existing benchmarks evaluate semantic query processing engines and data agents largely in isolation.
Benchmarks for semantic query processing engines, such as SemBench~\cite{sembench2026} and Bench-U~\cite{benchu2026}, execute fixed semantic queries, so they do not measure whether an agent can decide when a semantic operator is needed and how to specify it.
On the other hand, data agent benchmarks, such as DAB~\cite{dab2026} and KramaBench~\cite{kramabench2026}, focus on evaluating agents over structured or textual data without the use of these engines.

To address this gap, we introduce \emph{AvalancheBench}, a multimodal enterprise analytics benchmark for evaluating data agents that use semantic query processing engines as tools, rather than evaluating engines on fixed queries or agents without them.
Because representative enterprise data is rarely public, AvalancheBench generates observed data from a latent world specification and evaluates how accurately agents recover that world.
The benchmark contains three enterprise tasks (E-commerce, Travel, and Customer Support) with 58 questions spanning text, images, and documents.
We validate each task's latent world specification, generated data, questions, and reference answers for consistency and ecological validity.
Questions are provided together as a multi-part analytical brief,
reflecting report-oriented enterprise workflows. This leaves the
execution order to the agent and tests its ability to discover
dependencies, reuse computations, and maintain consistent answers. Agents must therefore
plan the analysis, decide which results to compute and reuse, and
answer interdependent questions consistently. This exposes the
\emph{avalanche effect}: an incorrect intermediate result can
propagate through legitimate dependencies, while inappropriate reuse
can create false dependencies and cause otherwise unrelated
questions to fail. Across the five configurations
evaluated in this paper, one execution of all three tasks costs
\$47 on average, ranging from \$31 to \$67.

Following the TPC tradition of reporting performance and
price-performance together~\cite{tpcc,tpch}, AvalancheBench evaluates systems using analytical fidelity and two
complementary cost measures. Monetary cost captures deployment cost
under current provider prices. We additionally introduce
\textit{capability-adjusted token cost} (\textit{CATC}), which weights token use by
model capability and measures how efficiently a system converts
capability-adjusted model resources into analytical fidelity. We
evaluate five model--harness configurations across Claude Code~\cite{claudecode}, Codex~\cite{codex}, and CoCo~\cite{coco},
using Snowflake Cortex AISQL~\cite{cortexaisql2026} as the semantic query
engine. In our evaluation, the best configuration reaches a macro analytical fidelity
of 0.82. Semantic operator calls account for 61\% of monetary cost
and 73\% of CATC, indicating that overall efficiency depends
substantially on how effectively an agent uses the engine.

Monetary cost and CATC lead to different deployment conclusions.
Among the 20 combinations of the five configurations with the three tasks and their macro average, 8 lie on both Pareto frontiers, 9 lie on neither, and 3 lie on exactly one.
One-sided membership distinguishes configurations that are
attractive because of current provider prices from those that use
capability-adjusted resources efficiently but are currently more
expensive. CATC therefore complements, rather than replaces,
monetary cost: monetary cost supports decisions under current prices,
whereas CATC offers a more stable, price-independent view of technical efficiency.

Our error analysis further reveals that agents fail primarily while
constructing the analysis, rather than because semantic operators
return incorrect outputs.
Across the four configurations analyzed in detail, 45.1\% of lost points come from failures in recovering facts from the latent world that require semantic computation.
Of all lost points, only 2.5\% are due to incorrect semantic operator outputs, whereas misusing or avoiding semantic operators together account for 36.3\%.
Requirement misinterpretation accounts for the largest share of lost points at 51.5\%.
Failures are also not independent: 29\% of lost points come from questions that fail by reusing another question's result, and root errors along with their
downstream consequences account for 42.2\% of all lost points.
These findings motivate explicit verification against the stated
requirements, better guidance on when and how to use semantic
operators, and additional verification of high-impact intermediate
results.

In summary, we make the following contributions:
\begin{enumerate}
    \item We introduce AvalancheBench, a multimodal enterprise
    analytics benchmark that evaluates how data agents plan and
    execute end-to-end analyses using semantic query engines,
    rather than merely executing fixed semantic queries.

    \item We develop a methodology for generating tasks from a latent world specification and validating them, which provides controlled ground truth while
    supporting realistic, interdependent analytical questions that expose the avalanche effect.

    \item We introduce CATC and evaluate analytical fidelity under both
    monetary and capability-adjusted cost, revealing when current
    provider pricing and underlying technical efficiency lead to
    different conclusions.

    \item We develop a trace-based failure taxonomy and
    dependency-aware analysis that identifies where errors originate,
    how they propagate, and which aspects of agent planning and
    semantic-operator use most need improvement.
\end{enumerate}

The remainder of the paper is organized as follows: 
Section~\ref{sec:avalanchebench} presents AvalancheBench and its
construction, Section~\ref{sec:task-design} describes the benchmark's
design goals, and Section~\ref{sec:metric-design} defines the
evaluation metrics. Section~\ref{sec:evaluation} presents the
experimental results and analysis, Section~\ref{sec:related} reviews
related work, and Section~\ref{sec:conclusions} concludes.

%% file: benchmark.tex
\section{AvalancheBench}
\label{sec:avalanchebench}

AvalancheBench casts agent-driven semantic analytics as latent world
recovery: using a semantic query engine, an agent must recover facts
and relationships from heterogeneous data to answer a multi-part
analytical brief. We next formalize the problem, describe its multimodal
realization, and introduce the three benchmark tasks.

\begin{figure*}[t]
  \centering
  \begingroup
  \small
  \emph{Travel Q3: Do polished ads get more clicks than natural-looking photos?}\par\medskip
  \begin{minipage}[t]{0.28\textwidth}
    \centering
    \textbf{Latent world $\world$}\par\medskip
    \setlength{\unitlength}{1pt}
    \begin{picture}(135,70)
      \put(67.5,58){\makebox(0,0){Subject}}
      \put(47,35){\makebox(0,0){\Large$\swarrow$}}
      \put(88,35){\makebox(0,0){\Large$\searrow$}}
      \put(25,12){\makebox(0,0){Format}}
      \put(110,12){\makebox(0,0){$\Pr(\text{click})$}}
    \end{picture}\par\smallskip
    No direct effect of format
  \end{minipage}\hfill
  \begin{minipage}[t]{0.04\textwidth}
    \centering\vspace{52pt}
    $\longrightarrow$
  \end{minipage}\hfill
  \begin{minipage}[t]{0.32\textwidth}
    \centering
    \textbf{Observed data $\data$}\par\medskip
    \includegraphics[width=0.48\linewidth]{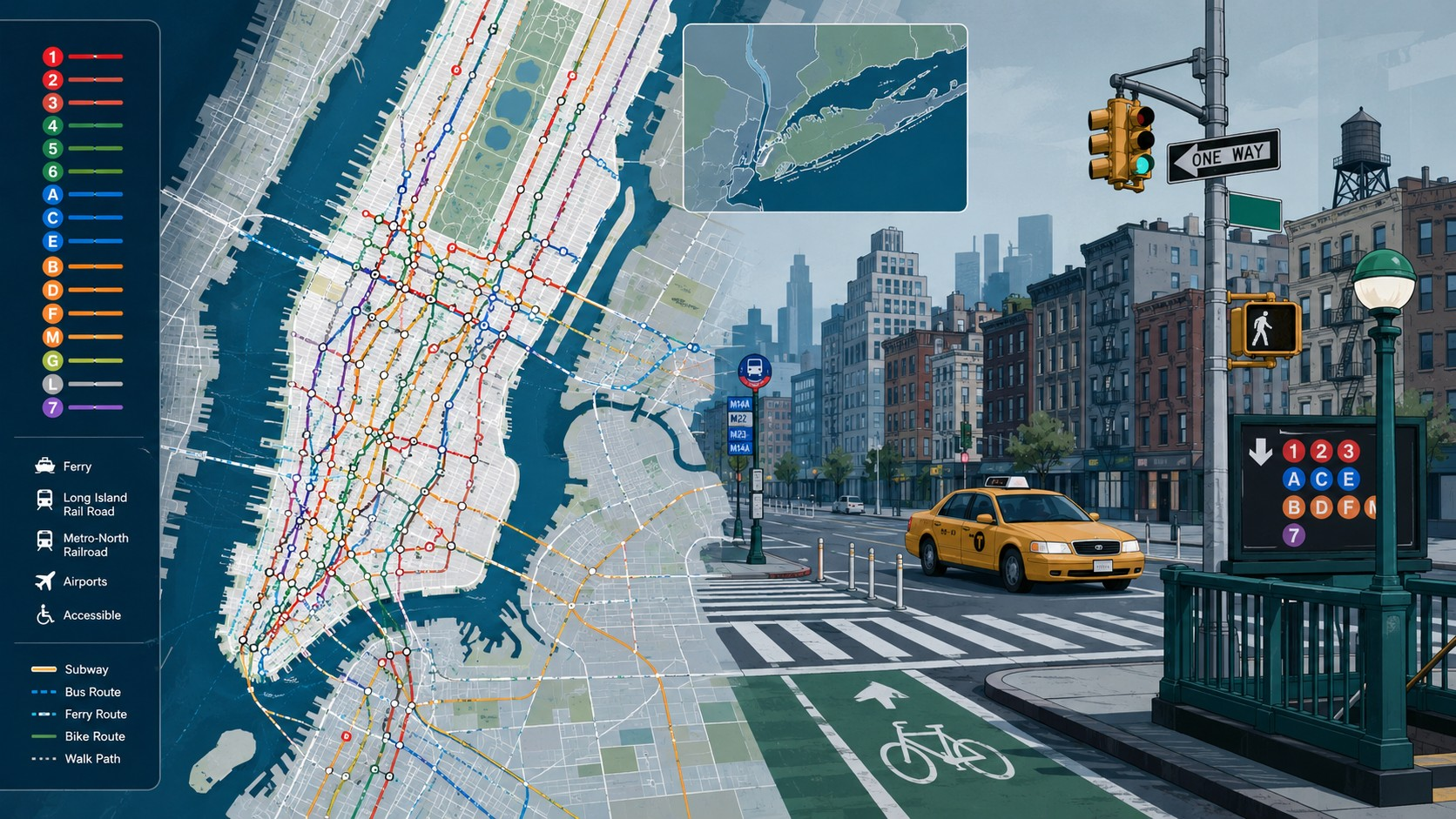}\hfill
    \includegraphics[width=0.48\linewidth]{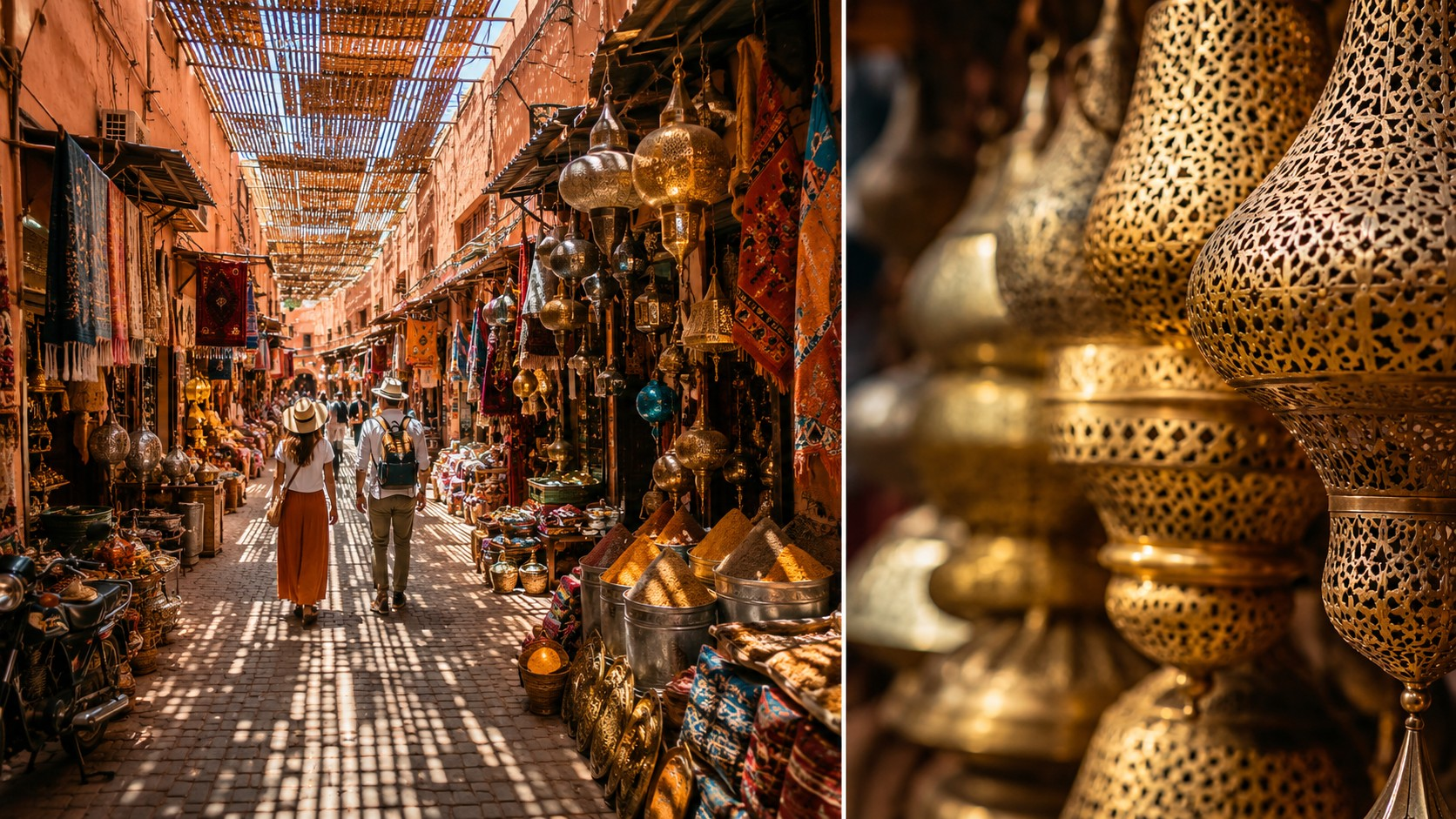}\par\smallskip
    \includegraphics[width=0.48\linewidth]{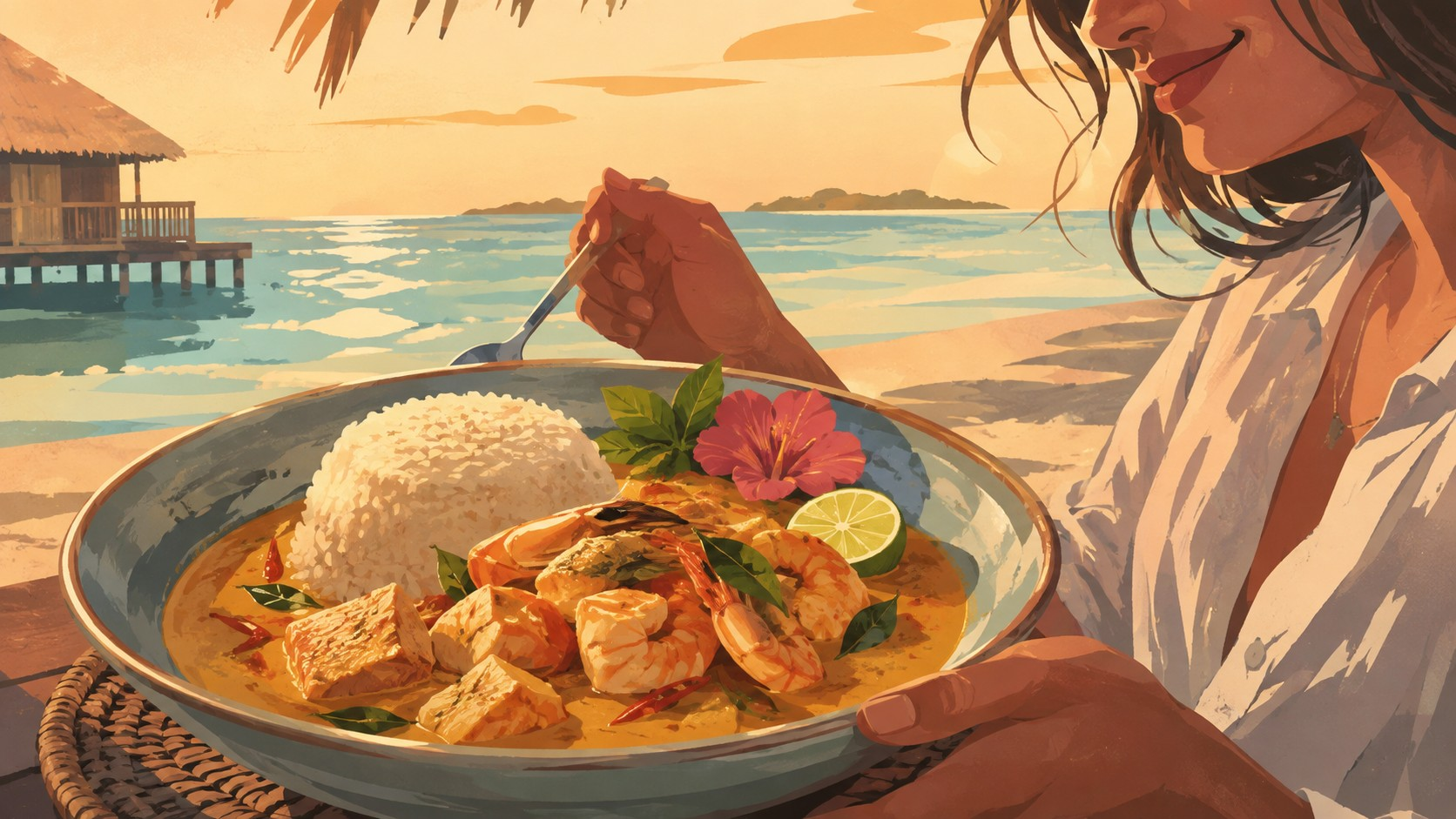}\hfill
    \includegraphics[width=0.48\linewidth]{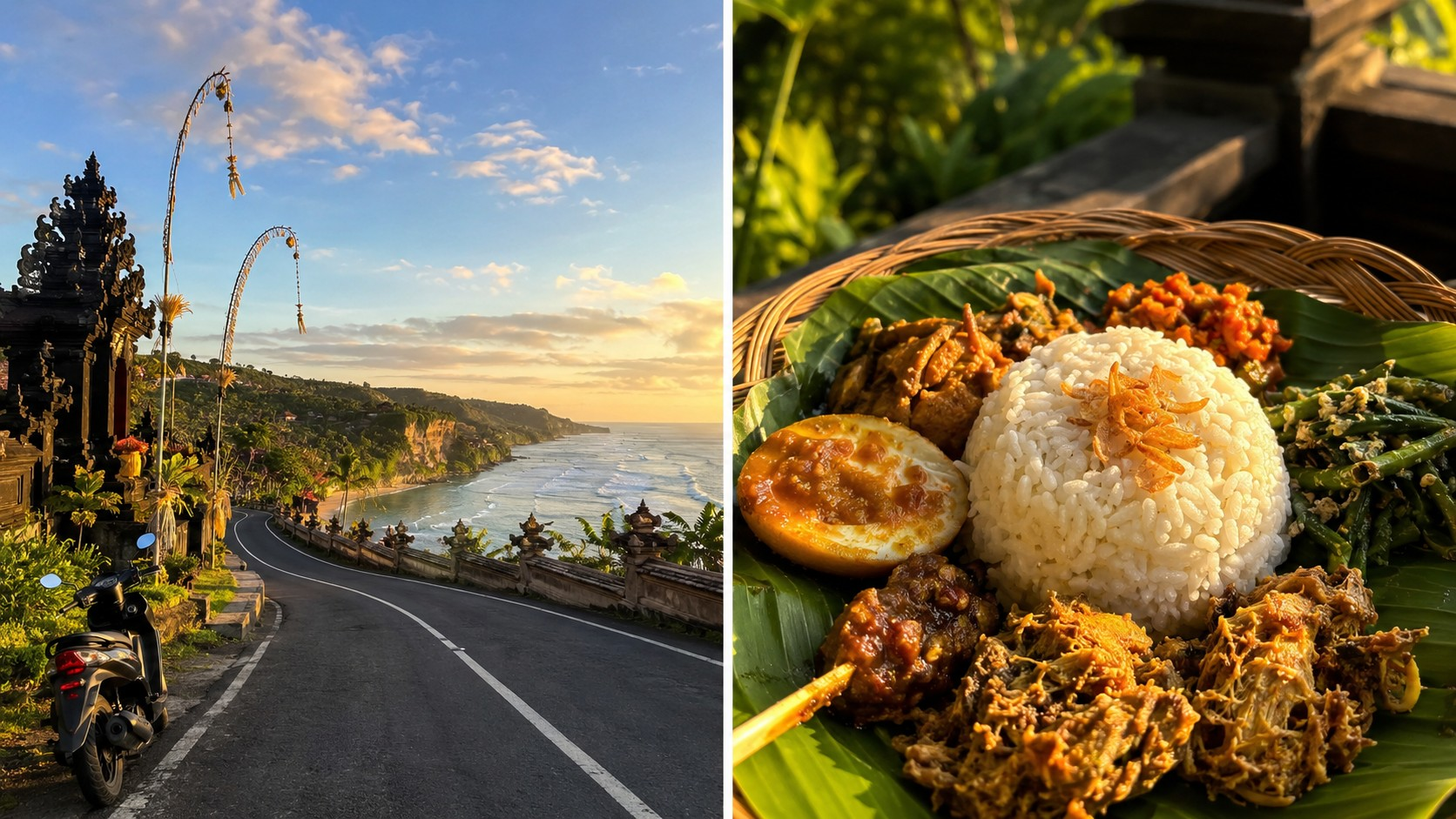}\par\smallskip
    Images + interaction records\\
    Subject/format not in tables
  \end{minipage}\hfill
  \begin{minipage}[t]{0.04\textwidth}
    \centering\vspace{52pt}
    $\longrightarrow$
  \end{minipage}\hfill
  \begin{minipage}[t]{0.28\textwidth}
    \centering
    \textbf{Recovery $\rightarrow\answers(\question)$}\par\medskip
    \begin{tabular}{@{}c@{}}
      Identify subject and format\\
      $\downarrow$\\
      Join on \texttt{image\_id}\\
      $\downarrow$\\
      Compare within subject groups
    \end{tabular}\par\medskip
    Relative click-rate gap\\
    $8.35\%\ \longrightarrow\ 3.88\%$\\
    Overall $\rightarrow$ same subject
  \end{minipage}
  \par
  \endgroup
  \caption{Benchmark mechanism illustrated by Travel Marketing: latent world $\world$ generates observed data $\data$; the agent answers $\question$ from $\data$ and is evaluated against reference answer $\answers(\question)$.}
  \Description{A left-to-right diagram. In the generator, image subject influences both format choice and click probability, but format has no direct effect on clicks; other click inputs are not shown. Four ad images are linked to interaction records, but their subject and format labels are not in the tables. The agent must identify those labels, join the records by image identifier, and compare formats within the same subject group. The relative click-rate gap falls from 8.35 percent overall to 3.88 percent when comparing within subject groups.}
  \label{fig:pipeline}
\end{figure*}

\textbf{Task Model.} We define an analytical task as a tuple $(\world, \data, \questions, \answers)$: a \emph{latent world} $\world$, the \emph{observed data} $\data$ generated from $\world$, a set of \emph{questions} $\questions$ about $\world$, and a map $\answers$ that assigns each question $\question \in \questions$ a reference \emph{answer} $\answers(\question)$ derived from $\world$.
Each reference answer comes with criteria for grading responses against that answer; together, a task's reference answers and grading criteria form its rubric.
Given only $\data$ and $\questions$, a system produces a report $\report$ with a response $\report(\question)$ to each question.
The system's goal is for each $\report(\question)$ to align with $\answers(\question)$, i.e., to \emph{recover} the relevant aspects of $\world$ from $\data$.

\textbf{Latent World Model.} AvalancheBench encodes each latent world in a specification file that defines its salient structure, such as persona distributions, temporal events, and product names.
From this file, we generate the observed data as a relational database, where each aspect is materialized either explicitly as structured values (e.g., numbers, categories, or dates) or implicitly in unstructured data (e.g., text, images, or documents).
The questions are given in a markdown file, and the system writes its report in markdown.
For synthesis and prioritization questions, the system also receives a published policy that makes the expected decisions unambiguous. 

\textbf{Multi-Question Analytical Brief.} We present all questions together as a single, multi-part analytical
brief. This models report-oriented workflows such as business
reviews, campaign assessments, and support audits, in which an
analyst receives several related questions and must produce a
coherent report. Unlike sequential benchmarks, which fix the request
order and restrict dependencies to earlier turns, AvalancheBench
leaves the execution order and reuse strategy to the agent. The
agent must discover dependencies and decide which intermediate
results to share, making reuse both a source of efficiency and
consistency, as well as a potential source of cascading failures.

\textbf{Scope of Discovery.} Discovery in AvalancheBench concerns latent analytical
structure (e.g., personas, visual concepts, recurring problems,
and cross-question dependencies), rather than locating unknown
tables or reverse-engineering opaque schemas. We therefore keep the
relational schemas manageable and documented so that failures can
be attributed to semantic recovery, operator selection, and
analytical planning rather than schema search or data integration.
Large, unfamiliar schemas are important enterprise challenges, but
including them here would confound the phenomena that AvalancheBench
is designed to isolate and diagnose.

\textbf{Motivating Example.}
In the Travel Marketing task (Figure~\ref{fig:pipeline}), $\world$ describes the advertised destinations, what each ad image shows (its \emph{subject}) and how it is presented (its \emph{format}), the audience's interests, and how likely visitors are to click, save an ad, or express interest in booking.
$\data$ contains the ad images and tables of campaigns, visitors, and interactions; an identifier links each image to the tables, but no table says what the image shows or how it is presented.
The questions ask which ads work, which audiences respond, and where to spend the budget.
Question $\question$ (Travel Q3) asks whether polished ads, such as brochure-style graphics, get more clicks than natural-looking photos.
In the generator, format has no direct effect on clicks, but polished formats are used more often for appealing subjects, so a naive comparison favors them.
The reference answer $\answers(\question)$ shows that the gap shrinks when comparing images of the same subject, such as food or a market.
The agent must therefore identify each image's subject and format, link the images to the interactions, and separate presentation from content.
We can vary the complexity by changing how often polished formats are paired with appealing subjects, without adding images or categories.
Because the same image labels feed audience and targeting analyses, one labeling mistake can affect several later answers: we call this the \emph{avalanche effect} (Sections~\ref{sec:complexity} and~\ref{sec:avalanche-effect}).

We now describe the three tasks (Table~\ref{tab:tasks}). For each, we detail its latent world, the data and questions the system receives, and what makes it hard (i.e., shortcuts that look reasonable but lead to a wrong answer).
We group the questions into types by what they require from the system (Table~\ref{tab:question-types}); every task mixes several types, and recommendation questions build on the answers to the others.
The tasks, including data, questions, and reference answers, are available in our repository at \url{\vldbavailabilityurl}.

\begin{table*}[t]
\centering
\small
\setlength{\tabcolsep}{5pt}
\caption{Overview of the three AvalancheBench tasks. \emph{Business rules} are documents given with the data, which some questions require applying: the investment policy ranks products for investment (defects first, then revenue at risk), and the knowledge base policy defines when a document is correct, partial, outdated, or conflicting, and how to rank fixes. \emph{Information to recover} is not explicitly materialized as a table column: the system must infer it from the data.}
\begin{tabular}{@{}p{2.6cm}p{4.3cm}p{4.9cm}p{4.9cm}@{}}
\toprule
& \textbf{E-commerce Reviews (Text)} & \textbf{Travel Marketing (Images)} & \textbf{Customer Support (Documents)} \\
\midrule
Tables & products, reviews, sales & destinations, campaigns, images, visitors, interactions, comments & tickets, ticket comments \\
Unstructured data & Reviews, half informative & Ad images; comments & PDF, DOCX, PPTX files; tickets \\
Business rules & Investment policy & --- & Knowledge base policy \\
Time span & 24 months & 12 months & 8 months, split into two periods \\
Information to recover & Review product; personas; event dates & Image content; visitor interests; events & Ticket problem; correct document; key facts \\
\bottomrule
\end{tabular}
\label{tab:tasks}
\end{table*}

\begin{table*}[t]
\centering
\small
\setlength{\tabcolsep}{5pt}
\caption{Question types, with one example each, for the E-commerce (E), Travel Marketing (T), and Customer Support (C) tasks.}
\begin{tabular}{@{}lll@{}}
\toprule
\textbf{Type} & \textbf{Example} & \textbf{Latent target} \\
\midrule
Extraction & At what widget count does a dashboard export stall? (C) & A fact shown only in a scanned table \\
Ranking & Rank all products by average rating. (E) & Product of each review, from its nickname \\
Segmentation & Which recurring problems drive support demand? (C) & Problem types behind noisy ticket labels \\
Comparison & Which persona--product pair is rated highest? (E) & Fishing customers' love of the down jacket \\
Temporal & Which monthly spikes come from a niche audience? (T) & External events and who responds to them \\
Issue detection & Which tickets cite the wrong document? (C) & Wrong citations and distractor documents \\
Recommendation & How would you split next quarter's budget? (T) & Visitor interests and their bookings \\
\bottomrule
\end{tabular}
\label{tab:question-types}
\end{table*}

\subsection{E-commerce Reviews (Text)}
\label{sec:task-ecommerce}

The E-commerce task simulates the analysis of customer reviews for a fictional apparel brand: which products satisfy which customers, when and why satisfaction changed, and where to invest.
No table records which product a review is about or which customer group (persona) its author belongs to; both can be inferred only from the review text.

\textbf{Latent World.}
The brand sells 20 products, from a cotton tank to a down jacket.
Its customers fall into five personas defined by their sport: golf, fishing, martial arts, rucking, and sailing.
Each persona has its own priorities (e.g., martial artists care about breathability and freedom of movement) and favorite products, some deliberately surprising.
For example, fishing customers love the down jacket for ice fishing, and sailors love the transit jacket for wind and spray.
Over 24 months, three events lower ratings: the cotton tank develops holes after washing (months 4 to 9), the sweat shorts develop holes (months 7 to 12), and a shipping delay affects every product (months 14 to 21).
Each review's rating is the product's base rating, shifted by the persona's preference and any ongoing event, plus small random noise.

\textbf{Observed Data and Questions.}
The system receives products, reviews, and sales tables; each review records a customer ID, a rating, a date, and its text.
Half of the reviews are written by an LLM given the product, persona, rating, and month: it uses a product nickname (e.g., ``NF puffer jacket'' for the NordForm Down Jacket), writes in the persona's voice, and mentions any ongoing defect or shipping delay.
The other half are short, generic reviews that carry no information (e.g., ``works as expected'').
The questions cover ratings, quality issues, personas, trends, and sales, and end with an investment ranking under the published policy, which puts product defects first.

\textbf{What Makes the Task Hard.}
Some nicknames are shared by several products (e.g., ``trail shoes'' for the men's and the women's model), so misattributed reviews break every product-level answer.
Personas must be found in the text and linked to sales through the customer identifier; generic retail segments give statistics for personas that do not exist.
Finally, the shipping delay starts while the sweat shorts' ratings are still recovering: blaming every rating drop on a product merges the two events and breaks the policy, under which a shipping problem is not a product defect.

\subsection{Travel Marketing (Images)}
\label{sec:task-travel}

The Travel task simulates the analysis of an ad campaign for a fictional travel agency: which ad images get engagement, from which visitors, and why.
No table stores what an image shows, so the system must look at the images.

\textbf{Latent World.}
The agency advertises eight destinations (e.g., Bali, Paris, Tokyo, Marrakech) on four channels: Instagram, Pinterest, email, and Google Display.
Each image has eight visual properties, such as its subject (e.g., a food close-up, a festival, a local market, a hotel room) and its format (e.g., a polished brochure shot or a casual photo).
Each visitor has a visible demographic profile (age, region, home city, device) and a hidden primary interest among six (e.g., food, festivals, luxury).
Each interest favors one image subject: a visitor whose main interest is food clicks on food close-ups about 2.6 times more often than on other images.
Clicks also depend on the channel, the season, and dated external events.

\textbf{Observed Data and Questions.}
The system receives the image files and six tables, which identify each image but say nothing about what it shows.
About half of the positive and neutral comments hint at their author's main interest (e.g., ``I would go just for the street food honestly'').
The questions cover audiences, image performance, trends, channels, niche audiences, and budget, phrased as a marketing team would ask them (e.g., ``Build the visitor segmentation you would actually use to run this account'').

\textbf{What Makes the Task Hard.}
First, format has no direct effect on clicks: polished shots were used more often for popular subjects such as festivals, so they look better overall, and comparing images of the same subject shrinks the gap from 8.35\% to 3.88\%.
Second, interests are independent of demographics, so grouping visitors by age, region, or device, as the tables suggest, cannot reveal what they respond to; visitors must be grouped by what they click on.

\subsection{Customer Support (Documents)}
\label{sec:task-support}

The Customer Support task simulates a review of the knowledge base of a fictional software company: find the most frequent customer problems, the document that solves each, the operational cost of missing or outdated documents, and what to fix first.
Some facts can only be read in scans, charts, diagrams, or screenshots.

\textbf{Latent World.}
Tickets are handled by three support teams, from the front-line service desk up to engineering, and the data spans two periods separated by a product release.
Customers face 14 recurring problems, each with its own symptoms, error codes, and frequency in each period.
For example, users get stuck in a loop when recovering their two-factor login: tickets talk about ``SMS fallback'' before the release and about a ``backup code loop'' after it.
Each problem has one key fact stored in a single document; for the login loop, it is which reset to run and how long the recovery window lasts, shown only in a diagram of an exception sheet.
The knowledge base also holds partial, outdated, and conflicting versions of the same guidance, and distractors about a similar product (12\% of the documents).
Problems with missing documentation are escalated to engineering much more often (48\% of tickets, versus 8\% for partially documented problems), are reopened more often, and take longer to solve.

\textbf{Observed Data and Questions.}
The system receives tickets and comments tables, the documents as PDF, DOCX, and PPTX files, and a published policy.
Tickets carry rough categories chosen by support agents but not the actual problem, refer to documents by approximate names (e.g., ``the old SMS recovery article''), and point to a wrong document 16\% of the time; there is no list of documents.
During generation, we check that each key fact shown in a scan, chart, diagram, or screenshot is absent from the document's text layer but recoverable with OCR.
The questions ask for the 14 key facts, cover ticket volumes, the effect of the release, and documentation gaps, and end by ranking fixes under the policy.

\textbf{What Makes the Task Hard.}
Treating the login loop as two problems because its wording changed, or merging it with a similar problem, breaks every later answer on volumes, gaps, and citations.
A document that looks relevant is not enough: the system must find the correct one and connect it to the right tickets.
Reading only the text layer misses the facts shown in images, and guessing a plausible value is wrong, not just incomplete.

%% file: design.tex
\section{Task Design}
\label{sec:task-design}

Key to synthetic data generation is the flexibility to define the complexity of the tasks we can evaluate.
That same flexibility, however, comes with the burden of ensuring that tasks are consistent and representative of the real world.
We therefore design tasks with three goals in mind: complexity (Section~\ref{sec:complexity}), consistency (Section~\ref{sec:consistency}), and ecological validity (Section~\ref{sec:ecological-validity}).

\subsection{Complexity}
\label{sec:complexity}

The \emph{complexity} of a task $(\world, \data, \questions, \answers)$ consists of \emph{world complexity}: the factors and
relationships in the latent world $\world$; \emph{observational complexity}: how those factors are represented in the observed data $
\data$; and \emph{question complexity}: the interpretation and analytical steps required to answer $\questions$. Domain experts set these dimensions when designing the latent world, data generator, and questions.

\textbf{World complexity} is determined by the latent world variables and their relationships. The complexity can be increased by overlapping effects, correlated
occurrences, or mixed memberships. In Travel, for example, ad images have
presentation format and image subject sampled dependently,
while subject influences engagement. A format may appear more effective because of the subject it shows---and a meaningful analysis needs to distinguish this by comparing formats within the same subject groups. 

\textbf{Observational complexity} concerns how the latent factors are represented in $\data$.
Relevant attributes may be expressed through variable language,
embedded in heterogeneous documents, or encoded visually rather
than exposed as structured fields. E-commerce requires interpreting
reviews to identify products and personas despite
surface variation. Customer Support requires recovering facts
from knowledge-base documents, including visually presented
evidence. Travel requires identifying image subjects and
presentation formats. Complexity is related to how difficult it is to recover answer-relevant information from the observations---which is influenced by modality, variability, noise, etc.

\textbf{Question complexity} concerns the difficulty of analytical operations to answer the questions and how explicitly the required interpretation is
specified. Questions range from retrieving a document fact to
linking recovered attributes with structured records, making
conditional comparisons, and synthesizing recommendations from
multiple findings. Their difficulty therefore depends on both
the analytical steps required and the guidance supplied about the
intended analysis.
Question complexity also arises across questions.
In real analyses, later conclusions build on earlier ones: once the personas or a set of recurring problems is established, the comparisons and recommendations that follow rest on them.
We say that a question depends on another when its reference answer builds on an intermediate result of that question.
Because all questions are posed at once, a system must itself identify these dependencies and decide which intermediate results to compute once and reuse.
Reuse avoids redundant work and keeps answers consistent with one another, but it exposes the system to what we call the \emph{avalanche effect}: a wrong shared result can spread to the questions that build on it, and reusing a result in a question that does not depend on it can make that question fail as well.
Section~\ref{sec:avalanche-effect} measures how many points systems lose to the avalanche effect.

The three dimensions of complexity provide a way to
construct new variants as systems improve. 
Our three tasks differ along each axis and span a wide range of difficulty (Section~\ref{sec:exp-results}).

\subsection{Consistency}
\label{sec:consistency}

A task $(\world, \data, \questions, \answers)$ is \emph{consistent} if it satisfies the following:
\begin{enumerate}
    \item \emph{Question--answer--world consistency}: each question $\question \in \questions$, read in the context of the task, has a unique answer under the latent world $\world$, given by $\answers(\question)$.
    That is, $\answers(\question)$ follows from $\world$, and $\world$ admits no other valid answer to $\question$.
    \item \emph{Observation--world consistency}: the observed generated data $\data$ faithfully realizes the latent world $\world$.
\end{enumerate}
Because each question has a unique answer under $\world$, given by $\answers(\question)$, and the data faithfully realizes $\world$, each $\answers(\question)$ is the unique correct answer on the generated data $\data$.
We operationalize consistency with checks for the two conditions above.

\subsubsection{Question--Answer--World Consistency}

To ensure that each reference answer $\answers(\question)$ follows from the latent world $\world$, we check that each question $\question$ can be correctly solved from the latent world specification file alone, where correctness is defined by $\answers(\question)$.
Then, to ensure $\answers(\question)$'s uniqueness, we check that each question $\question$ is specified well enough such that the latent world $\world$ does not admit multiple valid answers to $\question$.
We note that uniqueness is a stronger condition than consistency strictly requires: one could instead allow more open-ended questions paired with multiple valid answers. We adopt uniqueness because it is much easier to verify and leave open-ended questions for future work.
Uniqueness does not require each question to be fully specified on its own: a question may leave details implicit, as long as the latent world (as realized in the data), task description, other questions, and standard domain knowledge resolve them to a single answer.
This mirrors real analytical tasks, which are rarely fully specified (Section~\ref{sec:ecological-validity}).

As an example, to verify a question that asks about persona preferences, we can query structured metadata used to generate product reviews.
This metadata contains the ground-truth persona, product ID, and rating, so we can determine the correct answer with a simple group-by operation.

\subsubsection{Observation--World Consistency}
\label{sec:observation-world-consistency}

The main challenge in ensuring that the observed data $\data$ faithfully realizes the latent world $\world$ lies in the unstructured data. 
Structured attribute values are produced by code and are thus guaranteed to match the latent world specification, whereas unstructured values are generated by models and subject to errors.
Each unstructured item is generated by a model from the same fixed set of metadata, whose values are sampled from the distributions in the latent world specification.

To verify that a generated data item reflects its intended aspects, we sample observations and have domain experts check that they are consistent with the metadata used to generate it. For example, they may look at an informative product review and need to confirm that the ground-truth persona and product ID can be uniquely inferred from it. We use this feedback to harden the data generation process, and stop when domain experts confirm the consistency. 

\subsection{Ecological Validity}
\label{sec:ecological-validity}

A task $(\world, \data, \questions, \answers)$ is \emph{ecologically valid} if the latent world $\world$, the observed data $\data$, and the questions $\questions$ are representative of the real world.
We operationalize ecological validity along two fronts: the latent world $\world$ and questions $\questions$, which are authored, and the observed data $\data$, which is generated.

\subsubsection{World and Question Ecological Validity}

We employ domain experts to ensure that the latent world $\world$ and questions $\questions$ are ecologically valid (i.e., that they reflect the types of tasks and data observed in real companies). 
The table schemas and the set of possible aspect values are relatively easy to audit directly from the latent world specification file.
Whether the distributions defined over these aspect values are themselves realistic is less clear-cut.
For simplicity, we do not require experts to validate these distributions, since a system should ideally solve the task regardless of the underlying data distribution.

\subsubsection{Observation Ecological Validity}

Beyond faithfully realizing the latent world $\world$, the observed data $\data$ should be ecologically valid: generated data should resemble plausible data from the intended domain.
As with observation--world consistency, the difficulty lies in the unstructured data, which may contain generation artifacts that reveal its synthetic origin.
We assess the ecological validity of the generated data using the following procedure.

We assess the naturalness of generated data through domain-specific, sampled pairwise comparisons between our generated and human-created data. These checks cover, for example, e-commerce reviews or customer support documents. Each sampled generated datapoint is paired with a human-generated reference, with their presentation order randomized. A domain expert indicates which example is synthetically generated. We aggregate the judgments across the sample and iterate on the data generation process until manual review suggests that domain experts' ability to distinguish generated from human-created examples is close to chance level. 

\section{Metric Design}
\label{sec:metric-design}

To evaluate the performance of systems on AvalancheBench, we use several metrics to measure quality and cost.

\subsection{Analytical Fidelity}

Analytical fidelity measures how well a system's responses $\report(\question)$ align with the reference answers $\answers(\question)$ for each question $\question \in \questions$, according to the task's rubric.
The rubric defines four ordered grades for each question (zero, low, middle, and top), worth 0, 1, 2, and 5 points.

For a submitted report $\report$, a set $J$ of three LLM judges independently grades each response $\report(\question)$ against the reference answer $\answers(\question)$ and its grading criteria. Let $b_j(\report, \question) \in \{0, 1, 2, 5\}$ be the points awarded by judge $j \in J$. We define analytical fidelity as
\[
\mathrm{AF} :=
\frac{\sum_{\question\in\questions}
  \operatorname{median}_{j\in J} b_j(\report, \question)}
{5 |\questions|}
\in [0,1].
\]
Judges receive the matched report section when one can be identified;
otherwise, the evaluator falls back to the full report for that question.
An empty report receives zero points.

\subsection{Monetary Cost}

Following prior work~\cite{benchu2026,sembench2026,lotus2025,docetl2025,palimpzest2025}, we measure the monetary cost of each system.
We do not measure latency, since it depends on server load and configurations that vary across model providers and deployments.
For monetary cost, we price every model invocation during a system's execution, including those of the agent's model and of any model used in semantic queries, at public per-token list prices (Section~\ref{sec:exp-setup}).
We account for all token types, pricing input, cached input, cache write, and output tokens at their respective rates, and sum the resulting costs over all invocations.

\subsection{Capability-Adjusted Token Cost}

While evaluating monetary cost is helpful to understand the financial implications of a system's execution, we argue that it is insufficient to properly compare the benefits of cost optimizations implemented by different systems.
Monetary cost is subject to market fluctuations and financial incentives that affect provider token pricing.
Moreover, it conflates a system's own efficiency with the per-token efficiency of underlying LLMs: if a system adopts a newly released model that maintains the same capability but is cheaper per token, monetary cost improves even though the system's optimizations were not responsible for the improvement.
To address these limitations, we propose \emph{capability-adjusted token cost (CATC)} as a metric that isolates a system's cost from provider pricing and per-token efficiency gains of the underlying models.
We present one concrete instantiation of CATC; however, other formulations are possible, and we do not claim ours is uniquely correct.

Formally, let $\modelset$ be the set of all LLMs a system invokes during its execution, consisting of the agent's underlying model and any model used in semantic queries.
Further let $\invocations{\model}$ be the set of invocations to model $\model \in \modelset$.
Each invocation $\invocation \in \invocations{\model}$ is associated with a vector $\tokvec{\invocation} := (\toksin{\invocation}, \tokcin{\invocation}, \tokcwrite{\invocation}, \tokout{\invocation}) \in \N^4$ of input, cached input, cache write, and output token counts.
We instantiate the CATC of a system's execution as
\[
\text{CATC} := \sum_{\model \in \modelset} \sum_{\invocation \in \invocations{\model}} \capfactor{\model} \tokweights \cdot \tokvec{\invocation} \in \R_{\ge 0},
\]
where $\capfactor{\model} \in \R_{> 0}$ is the capability factor for model $\model$ and $\tokweights := (\tokinweight, \tokcinweight, \tokcwriteweight, \tokoutweight) \in \R_{\ge 0}^4$ is a vector of relative weights for input, cached input, cache write, and output tokens.
We describe how we compute $\capfactor{\model}$ in Section~\ref{sec:capability-factor}.
Our formulation treats model capability as a resource that a system consumes.
Intuitively, $\capfactor{\model} \tokweights$ captures the capability-adjusted cost of each token type for model $\model$, so that tokens spent on a more capable model are priced proportionally higher.
Taking the inner product $\capfactor{\model} \tokweights \cdot \tokvec{\invocation}$ then gives the capability-adjusted cost of invocation $\invocation$ based on its token counts $\tokvec{\invocation}$.
Without the capability factor, CATC would reduce to a weighted token count, under which a system could use more capable models at no cost.
For all models, we set $(\tokinweight, \tokcinweight, \tokcwriteweight, \tokoutweight) = (1, 0.1, 1.25, 5)$ following the ratios in Anthropic's pricing~\cite{anthropic_pricing}, which reflect the general ordering of serving costs for each token type~\cite{distserve2024,vllm2023}.
Note that in practice, the cost of processing a token grows with the length of the context it attends to, which our linear expression does not account for.
We nonetheless adopt the linear form for simplicity and consistency with providers' flat per-token pricing.
At a given level of analytical fidelity, a lower CATC indicates a more efficient system that processes fewer tokens or uses less capable models.
CATC does not, however, isolate a system's cost from differences in tokenizer or reasoning efficiency.


\subsubsection{The Capability Factor}
\label{sec:capability-factor}

We ground the capability factor $\capfactor{\model}$ in human-preference rankings from Arena~\cite{chatbotarena2024}, which fits a Bradley--Terry (BT) model~\cite{bradleyterry1952} to crowdsourced pairwise comparisons.
Each model $\model$ rated by Arena has a BT coefficient $\btcoef{\model}$, and the probability that $\model$ is preferred over another rated model $\model'$ is
\[
\Pr(\model \succ \model') = \frac{e^{\btcoef{\model}}}{e^{\btcoef{\model}} + e^{\btcoef{\model'}}}.
\]

Fixing a reference \emph{anchor} model $\anchor$, we define the capability factor of model $\model$ as the odds that $\model$ is preferred over $\anchor$:
\[
\capfactor{\model} := \frac{\Pr(\model \succ \anchor)}{\Pr(\anchor \succ \model)} = e^{\btcoef{\model} - \btcoef{\anchor}}.
\]
By construction, the anchor has $\capfactor{\anchor} = 1$, and the ratio $\capfactor{\model} / \capfactor{\model'}$ is itself the odds that $\model$ is preferred over $\model'$, which gives capability factors a multiplicative interpretation.
For example, a model that is twice as likely to be preferred over the anchor versus not has $\capfactor{\model} = 2$.
The choice of anchor is arbitrary: it simply defines the overall scale, but does not affect comparisons across systems.
We set the anchor to Llama 3.3 70B~\cite{llama2024}.

In practice, Arena reports its estimates of the BT coefficients as \emph{Arena scores} on the conventional Elo scale, $\arenascore{\model} = \frac{400}{\ln 10} \btcoef{\model} + C$.
Since $\capfactor{\model}$ depends only on the difference $\btcoef{\model} - \btcoef{\anchor}$, the offset $C$ cancels, and we compute it directly from reported scores as
\[
\capfactor{\model} = 10^{(\arenascore{\model} - \arenascore{\anchor}) / 400}.
\]

While Arena hosts different leaderboards, such as Text Arena~\cite{textarena}, Vision Arena~\cite{visionarena}, and Document Arena~\cite{documentarena}, we source all Arena scores from a single leaderboard so that every $\capfactor{\model}$ is expressed on a common scale relative to one anchor $\anchor$.
Scores from different leaderboards are not comparable, since each is a separate BT fit with its own offset and calibration.
Concretely, we use the overall Text Arena scores for each model, treating $\capfactor{\model}$ as a measure of general (rather than modality-specific) capability.
This ensures coverage of text-only models that may not appear on the Vision Arena or Document Arena leaderboards.
To limit confounding between capability and token usage, we take the style-controlled scores, which remove response length and formatting effects, and we source each model's score at the lowest reasoning effort listed on Arena, ideally none, so that reasoning effort is attributed to token counts rather than to the capability factor.

We acknowledge that human preference, as captured by Arena, should not be equated with capability, so $\capfactor{\model}$ is by no means a definitive measure of a model's capability.
Nonetheless, human preference correlates well with performance on capability benchmarks~\cite{arenaestimate2025,conditional2024}, so Arena scores serve as a reasonable proxy.

%% file: evaluation.tex
\section{Experimental Evaluation}
\label{sec:evaluation}

For our experimental evaluation, we first measure how well each configuration completes each task and what it spends to do so (Section~\ref{sec:exp-results}).
Because monetary cost and CATC can rank configurations differently, we then ask whether they lead to the same deployment choices (Section~\ref{sec:frontier}).
Finally, we explain where the lost points come from: we trace each failing question to its root cause (Section~\ref{sec:failure-modes}) and measure how far failures spread to other questions through the avalanche effect (Section~\ref{sec:avalanche-effect}).

\begin{table*}[t]
    \centering
    \footnotesize
    \setlength{\tabcolsep}{4pt}
    \newcommand{\best}[1]{{\boldmath$#1$}}
    \newcommand{\twoline}[2]{\textbf{\begin{tabular}[b]{@{}c@{}}#1\\#2\end{tabular}}}
    \caption{Analytical fidelity, monetary cost, and CATC per run (mean $\pm$ sample standard deviation over three runs). Bold marks the best mean per task among the configurations: highest fidelity, lowest cost.}
    \begin{tabular}{@{}lccccccccc@{}}
        \toprule
        & \multicolumn{3}{c}{\textbf{Analytical fidelity}}
        & \multicolumn{3}{c}{\textbf{Monetary cost (\$)}}
        & \multicolumn{3}{c}{\textbf{CATC (M)}} \\
        \cmidrule(lr){2-4}\cmidrule(lr){5-7}\cmidrule(lr){8-10}
        \textbf{Configuration}
        & \textbf{E-commerce} & \textbf{Travel} & \twoline{Customer}{Support}
        & \textbf{E-commerce} & \textbf{Travel} & \twoline{Customer}{Support}
        & \textbf{E-commerce} & \textbf{Travel} & \twoline{Customer}{Support} \\
        \midrule
        CoCo / GPT 5.6 Sol
        & \best{0.973\pm0.046} & \best{0.696\pm0.053} & $0.787\pm0.032$
        & \best{4.50\pm0.77} & $15.35\pm3.78$ & $34.98\pm3.94$
        & $13.97\pm8.02$ & $8.61\pm1.94$ & $26.04\pm2.16$ \\
        CoCo / Opus 4.8
        & $0.840\pm0.127$ & $0.552\pm0.122$ & $0.853\pm0.039$
        & $12.53\pm5.82$ & $10.57\pm1.16$ & \best{14.40\pm1.00}
        & \best{10.83\pm3.70} & $8.55\pm0.92$ & \best{8.05\pm0.60} \\
        CoCo / Sonnet 5
        & $0.818\pm0.076$ & $0.481\pm0.057$ & \best{0.861\pm0.032}
        & $4.97\pm0.45$ & \best{6.36\pm0.56} & $19.20\pm15.83$
        & $14.38\pm5.80$ & $10.40\pm1.40$ & $18.12\pm13.65$ \\
        Codex / GPT 5.6 Sol
        & $0.947\pm0.023$ & $0.585\pm0.122$ & $0.856\pm0.029$
        & $5.34\pm1.47$ & $7.42\pm0.88$ & $29.66\pm1.64$
        & $15.18\pm6.57$ & \best{5.58\pm1.20} & $23.07\pm1.19$ \\
        Claude Code / Opus 4.8
        & $0.836\pm0.008$ & $0.422\pm0.048$ & $0.755\pm0.041$
        & $23.94\pm32.54$ & $13.18\pm6.39$ & $30.27\pm20.40$
        & $25.25\pm20.68$ & $9.27\pm5.47$ & $22.36\pm16.79$ \\
        \bottomrule
    \end{tabular}
    \label{tab:exp-results}
\end{table*}

\begin{figure*}[t]
  \centering
  \includegraphics[width=\textwidth]{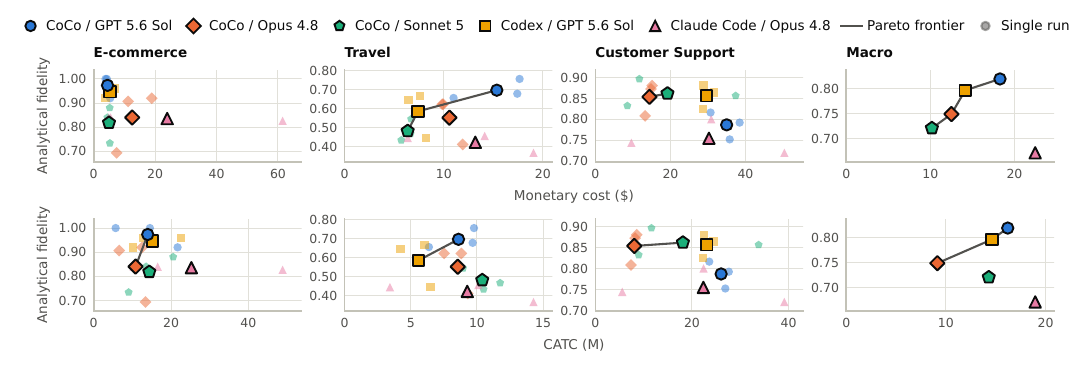}
  \caption{Analytical fidelity versus monetary cost (top) and CATC (bottom) per run; large solid markers are the mean of three runs and small faint markers are the individual runs. The line joins the configurations on the Pareto frontier of the means. Macro averages the three task means and shows no individual runs.}
  \Description{Eight scatter plots in two rows and four columns (E-commerce, Travel, Customer Support, Macro), with one point per configuration. The top row plots analytical fidelity against monetary cost in dollars and the bottom row against CATC in millions; a line joins the points on each panel's Pareto frontier. Faint markers show individual runs.}
  \label{fig:pareto}
\end{figure*}

\subsection{Setup}
\label{sec:exp-setup}
We evaluate the three AvalancheBench tasks described in
Section~\ref{sec:avalanchebench}, with three repetitions per task and
configuration. A configuration pairs an agent harness with an LLM; the resulting agent interactively
explores the evidence and invokes queries and other tools. We evaluate
CoCo~\cite{coco} with GPT 5.6 Sol~\cite{openai_gpt56}, Claude Opus 4.8~\cite{anthropic_opus48} and Claude Sonnet 5~\cite{anthropic_sonnet5}, Codex~\cite{codex} with
GPT 5.6 Sol, and Claude Code~\cite{claudecode} with Claude Opus 4.8. 
All agent configurations use Snowflake Cortex AISQL~\cite{cortexaisql2026} as the semantic query processing engine.
Structured data resides in Snowflake tables,
with images and documents available through stages. Each 
run has a two-hour execution budget.
The configurations are compared as complete systems:
their prompts, tools, state management, and execution
dates differ.
For monetary cost, we use the list prices for Anthropic~\cite{anthropic_pricing} and OpenAI~\cite{openai_pricing} models, and OpenRouter~\cite{openrouter_pricing} prices for open-weight models, as of October 29, 2026.
For CATC, we use the Text Arena~\cite{textarena} scores published on September 13, 2026 to compute each model's capability factor.

We use three LLM judges---Claude Opus 4.8, GPT 5.4~\cite{openai_gpt54}, and
Gemini 3.7 Flash~\cite{google_gemini37_flash}---at temperature zero. Each judge selects
one of four ordered rubric grades, and the committee awards the median
points. Across 45 reports and 870 question--report pairs,
the committee is unanimous on 67.0\% of pairs, with ordinal
Krippendorff's $\alpha=0.743$ computed on the ordered grades. Configuration-specific $\alpha$s range from 0.669 to 0.778.
Each judge separately preserves CoCo with GPT 5.6 Sol above
Codex on macro fidelity, although other macro rankings reverse.
These descriptive statistics establish inter-judge consistency; 
shared grading errors remain possible, and quantitative validation against independent human grading remains future work. 

\subsection{Main Results}
\label{sec:exp-results}

Table~\ref{tab:exp-results} reports analytical fidelity, monetary cost, and CATC per run for each configuration and task, as the mean and sample standard deviation over three runs.
Both cost measures cover the agent's own model calls and the semantic operator calls it issues through Cortex AISQL.
Figure~\ref{fig:pareto} plots fidelity against each cost measure for every task and for the macro average, which averages the three task means so that each task receives equal weight.

CoCo with Sol reaches the highest macro fidelity (0.819), followed by Codex with Sol (0.796), CoCo with Opus (0.748) and Sonnet (0.720), and Claude Code with Opus (0.671).
Travel is the hardest task for every configuration (0.422--0.696).
The macro ordering holds on Travel but not on Customer Support, where CoCo with Sonnet leads (0.861) and the macro leader drops to fourth (0.787).
With the model held fixed, the harness changes fidelity by up to 0.130 (CoCo versus Claude Code with Opus on Travel).

Configurations cost \$10.18--22.46 per run on macro average.
Semantic operator calls account for 61\% of all spending, and they vary far more across configurations than the agents' own model calls (\$4.23--17.36 vs.\ \$5.10--8.27 per run), so cost depends mainly on how agents use the engine.
With Opus held fixed, for example, Claude Code spends four times as much as CoCo on operator calls (\$17.36 vs.\ \$4.23).
Higher cost does not buy higher fidelity: Claude Code with Opus costs the most and has the lowest macro fidelity.

As with monetary cost, most of the CATC comes from semantic operator calls (73\%), but CATC ranks configurations differently from monetary cost: the cheapest configuration differs on E-commerce (CoCo with Sol under monetary cost, CoCo with Opus under CATC) and on Travel (CoCo with Sonnet under monetary cost, Codex with Sol under CATC).
One reason is that provider prices are not proportional to model capability.
For example, Sonnet's list price is 40\% of Opus's, but its capability factor is 93\% of Opus's (2.29 vs.\ 2.45), so Sonnet's tokens cost far less than Opus's in dollars but nearly as much in CATC.

\subsection{Frontier Agreement Between Cost Metrics}
\label{sec:frontier}

\begin{table}[t]
  \centering
  \footnotesize
  \setlength{\tabcolsep}{4pt}
  \caption{Frontier agreement under monetary cost (MC) and CATC. Each cell shows whether a configuration is Pareto-efficient under both cost measures, only one, or neither for each workload.}
  \begin{tabular}{@{}lcccc@{}}
    \toprule
    \textbf{Configuration} & \textbf{E-commerce} & \textbf{Travel} & \textbf{\begin{tabular}[b]{@{}c@{}}Customer\\Support\end{tabular}} & \textbf{Macro} \\
    \midrule
    CoCo / GPT 5.6 Sol     & Both      & Both     & Neither & Both     \\
    CoCo / Opus 4.8        & CATC only & Neither  & Both    & Both     \\
    CoCo / Sonnet 5        & Neither   & MC only & Both    & MC only \\
    Codex / GPT 5.6 Sol    & Neither   & Both     & Neither & Both     \\
    Claude Code / Opus 4.8 & Neither   & Neither  & Neither & Neither  \\
    \bottomrule
  \end{tabular}
  \label{tab:frontier}
\end{table}

Table~\ref{tab:frontier} summarizes how the choice of cost measure
changes the set of Pareto-efficient configurations in
Figure~\ref{fig:pareto}. The monetary frontier captures
cost--quality tradeoffs under today's provider prices, while the CATC
frontier reflects capability-adjusted resource use and is therefore
less tied to the current pricing schedule.
Of the 20 cells in Table~\ref{tab:frontier}, eight are on both frontiers, three on only one, and nine on neither.
Shared membership marks robust candidates for long-lived or ``sticky'' deployments,
where switching configurations incurs additional evaluation, tuning,
and integration costs. In contrast, a configuration absent from both
frontiers is dominated under each cost representation, although the
configuration dominating it may differ between the two.

The matrix also shows that this conclusion is workload- and harness-dependent. No configuration is jointly efficient for every workload: CoCo with Sol lies on both frontiers for E-commerce, Travel, and Macro, but not Customer Support, while CoCo with Opus is jointly efficient for Customer Support and Macro. Holding the model fixed does not preserve efficiency either. CoCo with Opus is CATC-efficient or jointly efficient on three workloads, whereas Claude Code with Opus appears on neither frontier for any workload; similarly, with Sol fixed, CoCo adds a jointly efficient E-commerce configuration that Codex does not. Thus, the results argue that the deployment choices should be made for the complete model--harness--workload combination rather than from a global model or harness ranking.

The three frontier disagreements are especially informative because they identify decisions that are conditional on the deployment environment. CoCo with Sonnet lies only on the monetary cost frontier for Travel and Macro: it is economically attractive at current prices, but alternative configurations achieve higher fidelity while consuming less CATC. For example, on Travel, Codex with Sol uses 46\% less CATC and achieves higher fidelity while costing only \$1.06 more. Such a choice may cease to be attractive as provider prices, hosting arrangements, or on-premise accounting change. Conversely, CoCo with Opus lies only on the CATC frontier for E-commerce: it uses less CATC than CoCo with Sol, but current pricing removes it from the monetary cost frontier. This configuration is technically efficient but not currently economical, and it may become more attractive under different pricing or deployment conditions.
CATC does not, however, replace deployment-specific cost accounting. Rather, the overlaps and disagreements reveal which recommendations are supported by both technical and economic efficiency, and which depend primarily on today's prices.

\subsection{Error Analysis}

We perform a detailed error analysis on four configurations that compare harnesses with the model held fixed (Claude Code and CoCo with Opus; Codex and CoCo with Sol) and models with the harness held fixed (CoCo with Sol and Opus), highlighting different failure modes and the avalanche effect.

\subsubsection{Failure Modes}
\label{sec:failure-modes}

\begin{table}[t]
  \centering
  \small
  \setlength{\tabcolsep}{3.5pt}
  \caption{Share of failing questions and points lost per failure mode, pooled over the 252 failures and 843 points lost in the four analyzed configurations and three runs. Bold rows sum to 100\%; indented rows break down the row above them.}
  \begin{tabular}{@{}lrr@{}}
    \toprule
    \textbf{Failure mode} & \textbf{\% of failures} & \textbf{\% of points lost} \\
    \midrule
    \textbf{Requirement interpretation} & \textbf{51.6} & \textbf{51.5} \\
    \addlinespace
    \textbf{Latent world recovery (LWR)} & \textbf{46.8} & \textbf{46.9} \\
    \quad Symbolic LWR & 2.0 & 1.8 \\
    \quad Semantic LWR & 44.8 & 45.1 \\
    \qquad Agent reading & 6.7 & 6.3 \\
    \qquad Symbolic proxy & 20.6 & 22.8 \\
    \qquad Semantic operator specification & 14.7 & 13.5 \\
    \qquad Semantic operator output & 2.8 & 2.5 \\
    \addlinespace
    \textbf{Result interpretation} & \textbf{1.6} & \textbf{1.7} \\
    \bottomrule
  \end{tabular}
  \label{tab:failure_taxonomy}
\end{table}

\begin{figure}[t]
  \centering
  \includegraphics[width=\columnwidth]{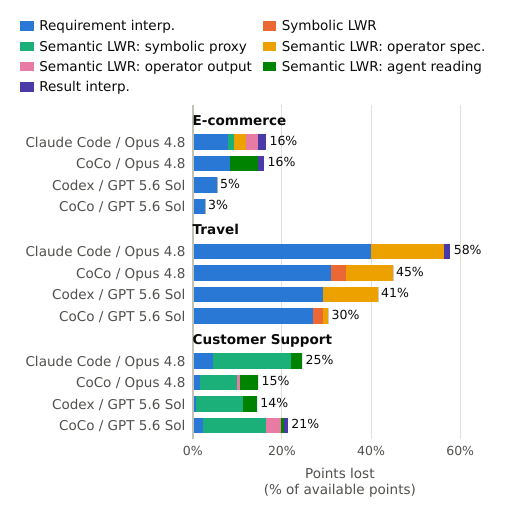}
  \caption{Points lost as a share of available points, by failure mode, task, and configuration (three runs per row).}
  \Description{Horizontal stacked bar chart with one bar per task and configuration, showing the percentage of available points lost, split into seven failure modes.}
  \label{fig:failure_taxonomy}
\end{figure}

To understand why questions fail, we assign each failing question in each run a single root cause based on the earliest analytical stage that went wrong.
We define a question as failing if it scores below full marks.
In general, answering a question requires the agent to (1) determine what the question requires, (2) recover the facts of the latent world that the answer depends on, and (3) draw conclusions from those facts.
This yields the following failure modes.
\begin{enumerate}
  \item \textbf{Requirement interpretation.} The agent misjudges what the question requires, such as what to compute or what to include.
  \item \textbf{Latent world recovery (LWR).} The agent understands the requirement, but a fact that the answer depends on is not recovered. We split LWR failures into two cases.
  \begin{itemize}
    \item \textbf{Symbolic LWR.} Recovering the fact fails at a step that requires only symbolic computation over the data (e.g., a wrong filter predicate on a numerical column).
    \item \textbf{Semantic LWR.} Recovering the fact fails at a step that requires semantic computation over the data. We further subdivide semantic LWR failures into four cases.
    \begin{itemize}
      \item \textbf{Agent reading.} The semantic computation is incorrect because the agent itself fails to read or misreads the relevant part of the content.
      \item \textbf{Symbolic proxy.} The semantic computation is incorrect because the agent uses a symbolic approach (e.g., regex) as a proxy for semantic computation.
      \item \textbf{Semantic operator specification.} The semantic computation is incorrect because the agent incorrectly uses semantic operators and therefore cannot recover the fact.
      \item \textbf{Semantic operator output.} The semantic computation is incorrect because the semantic operator outputs do not yield the correct end result despite the agent using the operators correctly.
    \end{itemize}
  \end{itemize}
  \item \textbf{Result interpretation.} The agent recovers the correct facts but fails to draw the correct conclusions from them.
\end{enumerate}
Table~\ref{tab:failure_taxonomy} pools failures and points lost by failure mode over the four configurations, and Figure~\ref{fig:failure_taxonomy} breaks them down by task and configuration.
We describe the key takeaways from these results.

\textbf{Agents rarely fail because semantic operators are wrong, but often because agents misuse or avoid them.}
Table~\ref{tab:failure_taxonomy} shows that semantic LWR accounts for 45.1\% of the points lost, but only 2.5\% come from incorrect semantic operator outputs.
For example, in Customer Support Q1, an \texttt{AI\_COMPLETE} call on a PDF document returned \texttt{renderer.export.hidden\_columns} even though the document reads \texttt{renderer.export\_hidden\_columns}, and the agent reported the wrong value in its answer.
Incorrect semantic operator specification costs 13.5\%, almost entirely in Travel and mostly through image groupings.
When Travel Q1 asks the agent to group the ad images, the rubric's top grade requires recovering at least five of the six subject families (i.e., groups of images by what they depict).
In one run, Claude Code with Opus classified the images with \texttt{AI\_COMPLETE} against a fixed list of twelve generic labels that it wrote after describing a single image, instead of describing all images and deriving the subject families from the descriptions with \texttt{AI\_AGG}.
This misuse of semantic operators missed some subject families.
Symbolic proxies cost the most among the semantic LWR failures at 22.8\%.
In Customer Support Q24, the question asks agents to ``resolve citations semantically using ticket context and document content'' when finding tickets that cite documents for the wrong problem.
In all four configurations, agents instead resolved citations by matching reference codes and keywords.
This missed the most common wrong citations, which point to documents about a similar product whose titles match the ticket's problem.
Better in-context guidance (e.g., skills) or training may help agents decide when and how to use semantic operators.

\textbf{Requirement interpretation accounts for the most points lost.}
As Table~\ref{tab:failure_taxonomy} shows, it is responsible for 51.5\% of the points lost.
For example, E-commerce Q6 asks the agent to compare sales in ``each affected 2024 peak period with the same calendar months in 2025''.
In one run, CoCo with Sol correctly identified the peak of the cotton tank defect as June--September in Q5 but compared sales over the defect's full affected window, April--September, in Q6.
In its other two runs, it compared the correct months.
Agents can thus meet such requirements but do not always do so.
Verifying each answer against the question's stated requirements could recover some of these points.

\textbf{Different tasks have different dominant failure modes.}
From Figure~\ref{fig:failure_taxonomy}, we see that requirement interpretation and semantic operator specification dominate in Travel, while symbolic proxies dominate in Customer Support.
In E-commerce, losses are small and mostly due to requirement interpretation.
The dominant failure mode of each task is the same in all four configurations.

\subsubsection{Avalanche Effect}
\label{sec:avalanche-effect}

\begin{table}[t]
  \centering
  \small
  \caption{Share of failing questions and points lost per dependency class, pooled over the 252 failures and 843 points lost in the four analyzed configurations and three runs.}
  \begin{tabular}{@{}lrr@{}}
    \toprule
    \textbf{Dependency class} & \textbf{\% of failures} & \textbf{\% of points lost} \\
    \midrule
    Independent & 59.9 & 57.8 \\
    Root & 13.5 & 13.3 \\
    True dependency & 26.2 & 28.5 \\
    False dependency & 0.4 & 0.5 \\
    \bottomrule
  \end{tabular}
  \label{tab:dependency_classes}
\end{table}

\begin{figure*}[t]
  \centering
  \includegraphics[width=\textwidth]{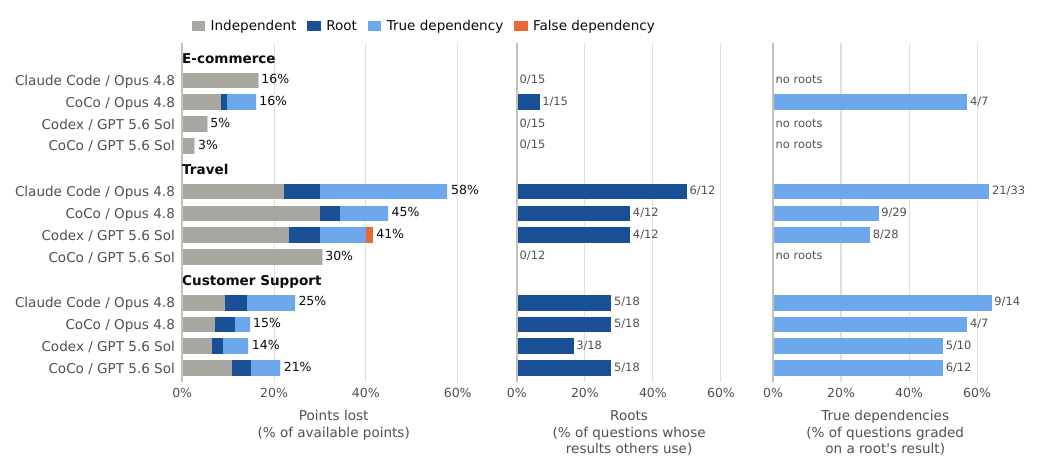}
  \caption{Avalanche effect by task and configuration (three runs per row). Left: points lost as a share of available points, split by dependency class. Middle: roots as a share of questions whose results others use; labels show roots / such questions. Right: true dependencies as a share of the questions graded on a root's result; labels show true dependencies / such questions, and ``no roots'' marks rows with no root to inherit from.}
  \Description{Three horizontal bar charts with one row per task and configuration, showing the percentage of available points lost split into four dependency classes, the share of questions whose results others use that fail as roots, and the share of questions graded on a root's result that fail as true dependencies.}
  \label{fig:avalanche_effect}
\end{figure*}

\begin{figure}[t]
  \centering
  \includegraphics[width=\columnwidth]{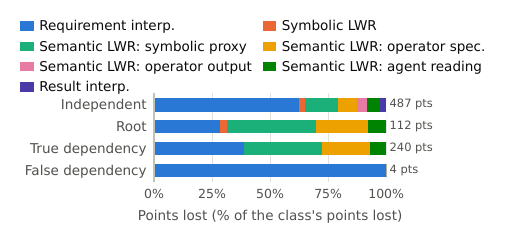}
  \caption{Share of each dependency class's points lost by failure mode, pooled over the four analyzed configurations and three runs. Labels give each class's points lost.}
  \Description{Horizontal stacked bar chart with one bar per dependency class, showing the share of the class's points lost from each of seven failure modes.}
  \label{fig:failure_mode_by_class}
\end{figure}

To measure the avalanche effect, we classify every failing question according to where its wrong result came from, using a dependency graph derived from each task's rubric.
We label each failing question with one of four dependency classes.
\begin{itemize}
  \item \textbf{Independent.} The question's wrong result is not inherited from a failing question, and no other question fails by reusing it.
  \item \textbf{Root.} The question's wrong result is not inherited from a failing question, and at least one other question fails by reusing it.
  \item \textbf{True dependency.} The question fails because it inherits a wrong result from a failing question on which it legitimately depends.
  \item \textbf{False dependency.} The question fails because it uses a result from a question on which it does not legitimately depend.
\end{itemize}
Table~\ref{tab:dependency_classes} pools failures and points lost by dependency class over the four configurations.
Figure~\ref{fig:avalanche_effect} breaks them down by task and configuration.
The left panel shows the percentage of available points lost by dependency class.
The middle panel shows how often shared results are wrong: among the questions whose results others use (five in E-commerce, four in Travel, and six in Customer Support), the fraction that fail as roots.
The right panel shows how far errors spread: among the questions graded on a root's result, the fraction that fail as true dependencies.
Figure~\ref{fig:failure_mode_by_class} breaks down the points lost in each dependency class by failure mode.
We highlight four main takeaways from these results.

\textbf{Errors that spread cost a large share of the points, and most of that cost is inherited.}
Table~\ref{tab:dependency_classes} shows that independent failures account for 57.8\% of the points lost, and roots and dependencies for the remaining 42.2\%.
True dependencies cost about twice as much as the roots that cause them (28.5\% versus 13.3\% of the points lost), and they are about twice as many (26.2\% versus 13.5\% of failures), so each root brings down about two further questions on average.
The image grouping described earlier illustrates how errors can spread.
In that run of Claude Code with Opus, Q1 recovered at most four of the six subject families and was a root.
Q5, Q16, Q17, and Q18 built interest cohorts (i.e., visitors who engage disproportionately with one subject family), budgets, and creative strategies on the same subject families.
Since these questions could not cover the missing families, they failed as true dependencies.
False dependencies, in contrast, are rare at 0.5\%: the only one occurred when Codex with Sol answered Travel Q16 with the home-city cohorts it built for Q6, even though the run had also built the interest cohorts on which Q16 legitimately depends.
Q6 is therefore a root, but it is not among the questions whose results others legitimately use, so it is the one root not counted in the middle panel.

\textbf{A root's error brings down many, but not all, of the questions graded on its result.}
From the right panel of Figure~\ref{fig:avalanche_effect}, we observe that between 29\% and 64\% of the questions graded on a root's result fail as true dependencies in Travel and Customer Support, and 4 of 7 in E-commerce, but no row reaches 100\%.
Questions graded on a root's result can still earn full marks for two reasons.
First, a question may reuse the root's result but not be graded on its wrong part.
For example, Customer Support Q20 asks which recurring problems lack approved documentation, and several runs listed two spurious problems alongside the two real ones, so Q20 failed.
Q23 reused this list but needs only the workarounds for the two real problems, so it earned full marks.
Second, the agent may not reuse the root's result at all.
For example, in one run of CoCo with Opus, Travel Q5 clustered visitors into broad segments such as high-intent visitors and low responders rather than interest cohorts, and failed as a root.
For Q8, Q14, and Q15, however, the run built separate interest cohorts, so all three earned full marks.

\textbf{Errors that spread come mostly from misusing or avoiding semantic operators.}
From Figure~\ref{fig:failure_mode_by_class}, we see that symbolic proxies and semantic operator specification make up most of the root bar, together accounting for 61\% of the points lost to roots.
For independent failures, they account for only 22\%, while requirement interpretation accounts for 63\%.
This is because the questions whose results others use typically require semantic computation over unstructured data, such as grouping the ad images in Travel Q1 or determining which recurring problems lack approved documentation in Customer Support Q20.
A misinterpreted requirement, in contrast, usually concerns a single question and stays local.
Since a true dependency usually fails in the same way as its root, symbolic proxies and semantic operator specification also account for 54\% of the points lost to true dependencies.
Spread nonetheless does not change which failure modes dominate, overall or per task.
Excluding true and false dependencies, requirement interpretation accounts for 56.3\% of the points lost, symbolic proxies for 18.5\%, and semantic operator specification for 10.9\%.

\textbf{Avoiding roots can matter more than avoiding independent failures.}
Figure~\ref{fig:avalanche_effect} shows that in Travel, CoCo with Sol loses more points to independent failures than Codex with the same model (30\% versus 23\% of the available points), but fewer points overall (30\% versus 41\%), because it has no roots.
The image grouping in Q1 highlights the difference.
Codex classified the images against a fixed list of labels, and Q1 was a root in two of three runs.
CoCo instead described each image with \texttt{AI\_COMPLETE} before deriving the subject families from the descriptions with \texttt{AI\_AGG}, and Q1 was never a root.
Agents may therefore benefit from allocating additional effort to results that many questions depend on.

%% file: related.tex
\section{Related Work}
\label{sec:related}

\textbf{Semantic query processing engines} like Snowflake's Cortex AISQL~\cite{cortexaisql2026}, DocETL~\cite{docetl2025,moar2026}, LOTUS~\cite{lotus2025}, and Palimpzest~\cite{palimpzest2025,abacus2026}, among many others~\cite{caesura2023,vectraflow2025,aryn2025,continuousprompts2025,stretto2026,samsara2025,docdb2025,aop2025,binder2023,databricks_ai_functions,duckdb_prompt,bigquery_ai_functions,uqe2024}, have been developed in recent years.
These engines feature a set of semantic operators, which augment traditional database operators with LLMs.
Novel query optimizers~\cite{abacus2026,moar2026} and efficient execution strategies~\cite{quest2025,bargain2025,semopkv2025,eleet2024,taskcascades2026,llmorderby2025,fdj2025} have been proposed to reduce the cost incurred by invoking LLMs, while benchmarks like SemBench~\cite{sembench2026} and Bench-U~\cite{benchu2026} evaluate these engines with respect to cost and quality.
Yet, these benchmarks only focus on evaluating engines on fixed semantic query workloads, rather than as tools used by agents for natural language tasks.

\textbf{Data agents} have substantially improved the automation of data work.
Agentic approaches achieve strong performance on earlier text-to-SQL benchmarks such as WikiSQL~\cite{wikisql2017}, BIRD~\cite{bird2024}, and Spider 2.0~\cite{spider2025}.
Yet, as exemplified by BEAVER~\cite{beaver2026}, data agents still struggle with text-to-SQL when faced with queries over real enterprise data that is private, messy, and idiosyncratic.
DAB~\cite{dab2026} and TACO~\cite{taco2026} similarly illustrate poor performance when answering questions over multiple databases, a scenario that is also common in enterprise analytics.
When tasked with constructing end-to-end ELT pipelines, ELT-Bench~\cite{eltbench2025} also demonstrates much room for agent improvement.
To scope our work, AvalancheBench focuses on analytical complexity arising primarily from unstructured data modalities, rather than the complexities described above.
TAG-Bench~\cite{tag2025} proposes translating natural language questions into semantic queries using an LLM to accomplish a broader range of tasks beyond text-to-SQL and RAG, but the semantic queries it evaluates are hand-written.
KramaBench~\cite{kramabench2026} tests agents on data-to-insight pipelines over data lakes of structured and unstructured textual data, but does not focus on evaluating agents with semantic query processing engines.

\textbf{Synthetic benchmarks} provide finer control over task complexity than benchmarks constructed from fixed real-world datasets.
These benchmarks can contain synthesized questions~\cite{sqlstorm2025,beaver2026,benchevolver2026,taco2026}, data~\cite{taubench2025,gaia22026}, or both~\cite{crmarena2025,crmarenapro2026,enterprisebench2025,appworld2024}.
Synthesized questions can include LLM-generated SQL queries~\cite{sqlstorm2025,beaver2026}, coding problems~\cite{benchevolver2026}, or natural language questions~\cite{taco2026}.
Synthesized data can be structured or unstructured.
Structured data can be sampled procedurally~\cite{taubench2025}, whereas unstructured data are often produced using generative models~\cite{crmarena2025,crmarenapro2026,gaia22026}.
Most relevant to our work are CRMArena~\cite{crmarena2025} and CRMArena-Pro~\cite{crmarenapro2026}, which synthesize CRM databases from latent world specifications.
Unlike AvalancheBench, however, these benchmarks do not evaluate agents equipped with semantic query processing engines and lack multimodal tasks.

\textbf{Error propagation} occurs when an error in an earlier result carries over to later results that build on it.
Data analysis benchmarks that pose sequences of dependent requests, such as DSEval~\cite{dseval2024}, Tapilot-Crossing~\cite{tapilot2024}, and LongDS~\cite{longdsbench2026}, differ in whether and how they measure it.
Tapilot-Crossing rules it out by evaluating each turn given a curated, error-free interaction history.
DSEval by default restores the correct results of earlier problems before each problem, and measures propagation only in a separate setting, as an aggregate drop in pass rate.
LongDS lets errors propagate and labels failed turns caused by errors from earlier turns.
In all three, however, requests arrive one at a time, so the order of the analysis is fixed by the request sequence and each request can build only on earlier ones.
In contrast, AvalancheBench poses all questions at once, so a system must plan the analysis itself, including which intermediate results to compute and reuse.
We classify each failing question by whether it originates an error or fails through the avalanche effect, either inheriting an error through a true dependency or reusing a result through a false dependency, and quantify the points lost to each.

%% file: conclusions.tex
\section{Conclusions}
\label{sec:conclusions}

AvalancheBench evaluates data agents not only by whether they
produce correct answers, but also by how they construct, reuse, and
propagate semantic results. Across three multimodal enterprise
analytics tasks, the best model--harness configuration reaches a
macro analytical fidelity of 0.82, but no configuration is uniformly
best across workloads or cost measures. Monetary cost and CATC
produce different Pareto frontiers, showing that efficiency is a
property of the model--harness--workload combination and that
configurations favored by current provider prices are not always
those that use capability-adjusted model resources most efficiently.

The failure analysis reveals two central reliability challenges.
First, agents fail primarily in constructing the analysis:
requirement interpretation accounts for 51.5\% of lost points, while 
misusing or avoiding semantic operators costs 36.3\%.
Incorrect operator outputs cost only 2.5\%. 
The main bottleneck is therefore not operator reliability, but how agents interpret requirements, determine the need for semantic computations, and specify them correctly.
Second, failures are not independent.
Questions that fail by reusing another question's result account
for 29\% of lost points;
including the root errors behind them, avalanches involve
42.2\% of all lost points. A relatively small number of incorrect
shared results can cause disproportionate downstream
damage.

These findings suggest actionable improvements for building reliable data agents. Agents need verification of their answers against the stated requirements, better guidance on when and how to
use semantic operators, and dependency-aware analysis that
allocates additional effort to high-impact intermediate results.
They should also track provenance and contain or recompute erroneous
results before they spread. By making these behaviors measurable,
AvalancheBench provides a foundation for evaluating progress toward
reliable and efficient semantic data analysis with agents.

%% file: acks.tex
\begin{acks}
We thank the Snowflake CoCo and AISQL teams for their support.
This material is based upon work supported by the National Science Foundation Graduate Research Fellowship Program under Grant No.~2439559. Any opinions, findings, and conclusions or recommendations expressed in this material are those of the authors and do not necessarily reflect the views of the National Science Foundation. 
\end{acks}